\documentclass[twocolumn,preprintnumbers,amsmath,amssymb]{revtex4}
\usepackage{graphicx}
\usepackage{dcolumn}
\usepackage{bm}
\begin{document}
\title{\bf Van der Waals energy surface of carbon nanotubes sheet}
\author{S. Motahari$^{1}$, F. Shayeganfar$^{2}$ and  M. Neek-Amal$^{1}$\\
\small $^1$Department of Physics, Shahid Rajaee Teacher Training
University, Lavizan,
Tehran 16788, Iran.\\
\small $^2$ Department of Physics, Sharif University of Technology,
Tehran P.O.Box 1155-9161, Iran.}
\begin{abstract} The van der Walls interaction between a carbon nanotubes sheet (CNTS)
 and a rare gas atom,  is studied using both atomistic and continuum approaches.
 We present analytical expressions for the van der Waals energy of continuous nanotubes
 interacting with a rare gas atom. It is found that the continuum approach dose not properly
  treat the effect of atomistic configurations on the energy surfaces.
 The energy barriers are small as
compared to the thermal energy, which implies the free motion above the CNTS in heights
about one nanometer.  In contrast to the energy surface of a graphene
sheet, the honeycomb lattice structure in the energy surface of a
CNTS is imperceivable. Defects alter the energy surface which
therefore influence the gas absorbtion mechanism.
\end{abstract}
\maketitle
\section{Introduction}
In the past two decades, various properties of carbon nanotubes such
as their unique structure and mechanical, chemical and electronic
properties, have been attracted much attention~\cite{cntBOOK}. Zhang \emph{et al} used
a solid state sheet fabrication technique and could synthesize highly oriented free standing carbon nanotubes
sheets (CNTS)~\cite{science2005}. The produced CNTS combines high
transparency with high electronic conductivity and are highly
flexible and provide giant gravimetric surface areas. The
long, high perfection nanotubes are needed to maximize the electrical and thermal
conductivities and mechanical properties~\cite{science2005}. Recently Li \emph{et al}~\cite{Li2010}
have fabricated highly engineered ultra-thin anisotropic single wall
carbon nanotube films formed on three dimensional micro-patterned
substrates. Therefore CNTS can be considered as a real multifunction
sheet for the future technologies.

The two-dimensional structure of CNTS makes it possible to examine
the possibility of the two-dimensional motion of an external particle
 over them~\cite{PRL2003,nanotechnology2002}.
The absorbtion of rare gas on the CNTS can change the lattice
parameter of nanotubes. In this regard Bienfait~\emph{et al} found experimentally
the dilation of the apparent carbon nanotube bundle in the presence
of argon gas atoms~\cite{PRL2003}.

 The potential energy surface (ES) of CNTS
explains different motion-related phenomena at nanoscale as well as
the various directed motions on the carbon nanotube-based
motors~\cite{neek, science2008}. Since rare gas atoms are
experimentally used to control temperature, pressure and other
thermodynamical properties of nanoscale
materials~\cite{elias,nature}, it is important to investigate the
van der Waals (vdW) potential energy surface around a CNTS.
Recently, we found a diffusive motion for the motion of $C_{60}$
molecule over graphene sheet and depicted that the
energy barriers are small as compared to the thermal energy~\cite{neek}. By
employing a different approach, i.e. Langevin dynamics, we showed
that a rare gas over graphene sheet has a diffusive
motion~\cite{Lajavardipour}.

There are two common approaches for studying the various properties
of carbon nanotubes based nanostructures: 1) atomistic and 2)
continuum. Yakobson \emph{et al} \cite{yakobson} have shown
that the mechanics of carbon nanotubes is interpretable using the
continuum elasticity of shells. The pioneering work of Yakobson and co-workers
made common practice of using of continuum models for simulations of
mechanical properties of nanotubes~\cite{euroconf,prb2005,apl2009}.
In recent studies, it has been shown that atomistic and continuum approaches
can be combined together to study mechanical behavior of different materials~\cite{euroconf}.
However, in the present study, we will show that employing of the continuum approach for energy based properties
(but not elastic energy) is not appropriate.

In this study the ES of a CNTS due to the vdW interaction with a
rare gas atom is investigated. We show that the ESs are periodic in
the planes which are parallel to the CNTS layers. We find that
there are significant differences between ES of the atomistic model
and those predicted from the continuum model. The binding energies
and the physical-bond distances obtained from the atomistic model
 differ from the continuum model. We conclude that the
continuum model is inappropriate approach for energetic
consideration and yields inaccurate results  for the chemical absorbtion of
atoms into the CNTS. However mechanics of atomistic systems
 (carbon nanotubes) greatly benefits
from continuum mechanics~\cite{euroconf,yakobson}. In addition,
 we have investigated the ES of graphene and compared it to the one obtained for CNTS.
  It is found that the honeycomb lattice feature, which is observed in the ES of graphene,
   does not appear in the ESs of CNTS.
 This is attributed to the large curvature of carbon nanotubes in the CNTS sheet

Results for the CNTS comprises the zigzag nanotubes differ from the CNTS
made of other types of nanotube chiralities.
From our barrier energy calculations, it is found that the barriers for the motion of
the rare gas atom on a CNTS become shallow at positions away from the CNTS.
 Moreover, the components of the
force are calculated and are compared to those obtained from the
continuum model. We show that the simplest kind of defect (i.e.
vacancy), destroys the periodicity of the ES and reduces the energy
barriers.

This paper is organized as follows. In Sec. II and III the details of
atomistic and continuum models are presented
Sec.~IV contains main results including those for the energy
surfaces, force calculation, numerical solution for the equation of
motion and the results for ES of the defected CNTS. In Sec. V the results are summarized.

\begin{figure*}
\begin{center}
\includegraphics[width=0.4\linewidth]{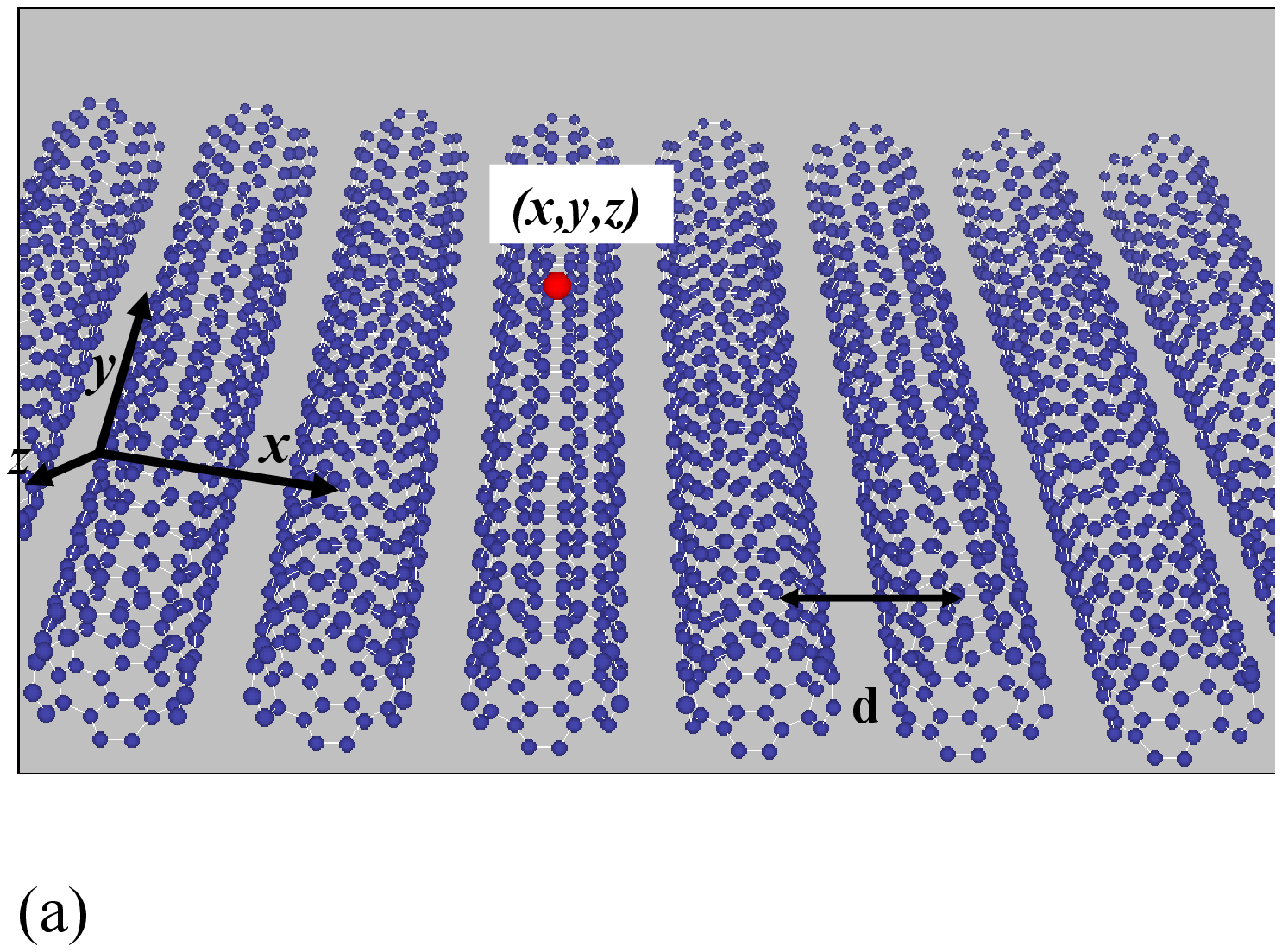}
\includegraphics[width=0.5\linewidth]{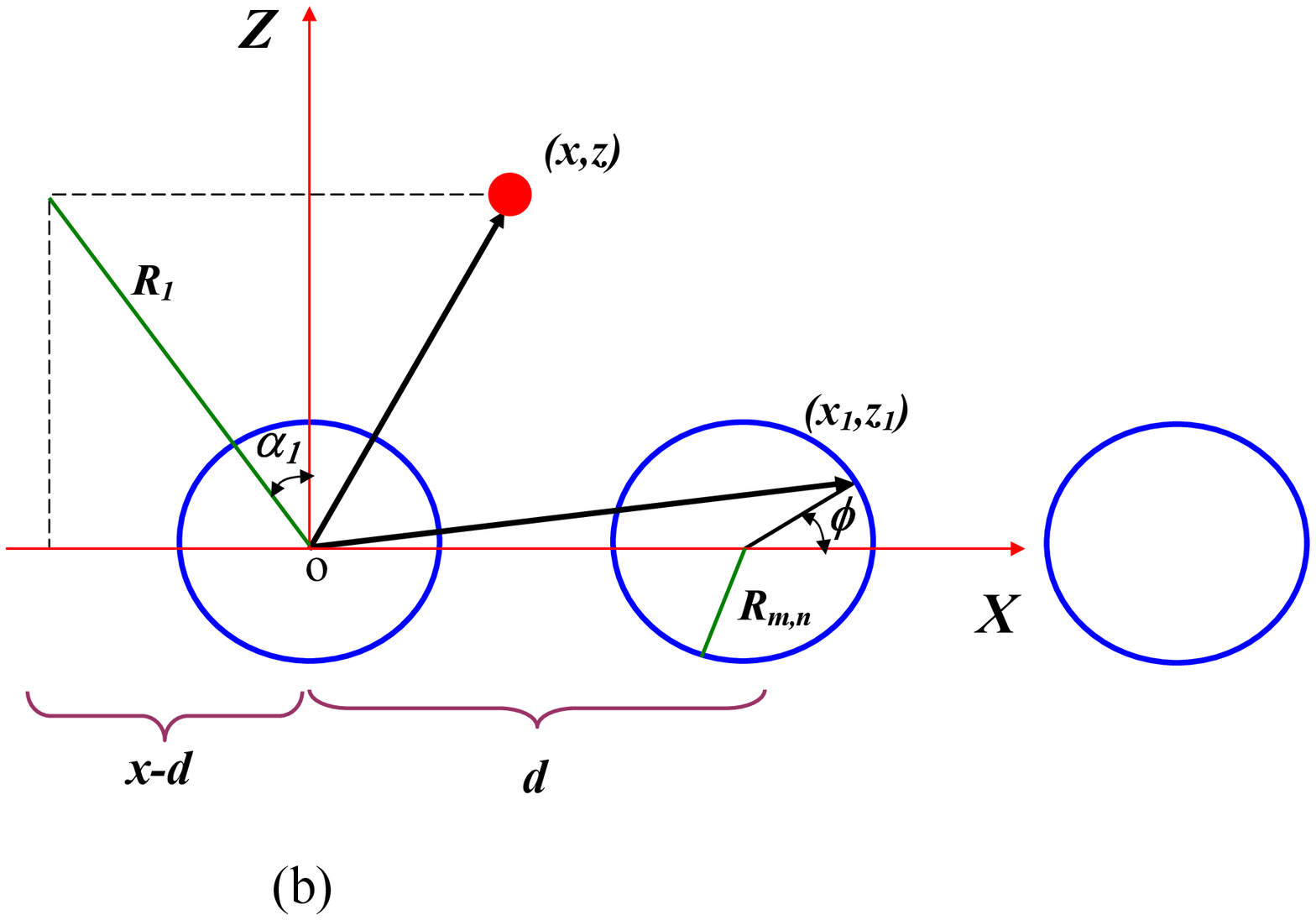}
\caption{(Color online)  (a) The atomistic model of a CNTS.
 The top layer of an AC-CNTS comprises eight nanotubes with $M$\,=\,400  and $d$\,=\,10\,\AA.
~(b) The continuum model of a CNTS. The front view of three continuous tubes with the relevant variables for
Eq.~(\ref{Eq_i}). The red sphere is an argon atom located at $(x,y,z)$
position. \label{figmodel} }
\end{center}
\end{figure*}

\section{The atomistic model}

 A CNTS comprises many layers which are stacked on
top of each other and are along $z$-axis. Many aligned carbon
nanotubes which are parallel to the $y$-axis construct each layer.
The carbon-carbon bond length, $a_0$, is 1.42\,\AA. In our model, the origin of the
xyz Cartesian coordinate system is set to be at the center of the top layer. Carbon nanotubes
form different types, which can be described by the
chiral vector with two indexes $(m,n)$, where $m$ and $n$ are
integers of the chiral vector equation $\overrightarrow{R}\,=\,
m\overrightarrow{a}_1 + n\overrightarrow{a}_2$. Here
$\overrightarrow{a}_1$ and $\overrightarrow{a}_2$ are two basic
vectors of graphene lattice and are given by
$\overrightarrow{a}_1\,=\sqrt{3}a_0\hat{i}$ and
$\overrightarrow{a}_2\,=\sqrt{3}/2a_0\hat{i}+3/2a_0\hat{j}$,
respectively. If $m$\,=\,0, the nanotubes are called zigzag (ZZ). If
$n\,=\,m$, the nanotubes are called armchair (AC). Otherwise, they are
called chiral (CR). The radius of a carbon nanotube with $(m,n)$
indexes is given by
$R_{m,n}\,=\,\frac{\sqrt{3}a_0}{2\pi}\sqrt{m^2+n^2+mn}$.

Figure~\ref{figmodel}(a) depicts eight aligned (5,5) carbon
nanotubes from the top layer. Graphene is a planar structure made of
carbon with one-atom thickness, where each carbon
 atoms have sp$^2$ hybridization with their neighboring atoms and are densely
  packed in a honeycomb crystal lattice. Carbon nanotubes can be formed by rolling up graphene sheets.
  In order to study the  van der Waals interaction
between an argon atom and a CNTS, we have employed the Lennard-Jones
(LJ) potential. The LJ potential provides both the repulsive and
attractive nature of the interaction between an uncharged noble gas
and each atom of the CNTS. The LJ potential is the widely used
potential in various simulations~\cite{prb2010} for two interacting
uncharged particles, i.e.
$u(r)\,=\,4\varepsilon((\sigma/r)^{12}-(\sigma/r)^6)$, where $r$ is the
distance between two atoms, $\varepsilon$ is the depth of the
potential well, and $\sigma$ is the distance at which the potential
becomes zero. To model the interaction between two different types
of atoms such as carbon and argon, we adjust LJ parameters using the
equations $\varepsilon\,=\, \sqrt{\varepsilon_1 \varepsilon_2}$ and $
\sigma\,=\,\frac{\sigma_{1}+\sigma_2}{2}$. We use the parameters of
carbon and argon atoms as $\sigma_1\,=\,3.369\,$\AA$,~
\varepsilon_1\,=\,2.63$\,meV, $\sigma_2\,=\,3.405$\,\AA~and
$\varepsilon_2\,=\,10.23$\,meV, respectively.

Total potential energy stored between an argon atom and a CNTS is written as the sum
over all contributions of the carbon nanotubes in each layer,
\begin{equation}\label{Eq1}
E_T(x,y,z)\,=\,4\varepsilon{\sum^{N/2}_{i\,=\,-N/2}}{\sum^K_{t=0}}{\sum^M_{j=1}}{\sum^2_{l=1}}
(-1)^{l+1}(\frac{\sigma}{\delta r})^{12/l},
\end{equation}
where $\delta r\,=\,|\overrightarrow{\bf{r}}
-(\overrightarrow{\bf{r}}^i_{j,t}-t\,D\,\widehat{\emph{k}})|$ and
$\overrightarrow{\bf{r}}\,=\,x\,\hat{\emph{i}}+y\,\hat{\emph{j}}+z\,\hat{\emph{k}}$
refer to the position of the argon atom and
$\overrightarrow{\bf{r}}^i_{j,t}$ is the position vector of $j^{th}$
carbon atom in the $i^{th}$ carbon nanotube (of the $t^{th}$ layer).
Each nanotube has $M$ atoms and the total number of carbon nanotubes in each layer is
$N+1$. The sum over $l$ also, is written for varying of the powers
between $12$ or $6$. The sum over $t$ counts the number of layers
($K$ is the total number of layers) in different height along
$z$-axis, such that each layer is separated by $D$ from its neighboring layer.
 Note that each layer is shifted along the $x$-axis with respect to
 its top and bottom neighbors,
 i.e. $\overrightarrow{\bf{r}}^i_{j,t\pm1}\,=\,\overrightarrow{\bf{r}}^i_{j,t}\pm\frac{d}{2}\widehat{i}$
 where $d$ is the lateral distance between tubes in each layer, see
Fig.~\ref{figmodel}. In the CNTS having many layers,
we show that the main contribution of the energy comes from the
top layer, thus setting $K=1$ would be a good approximation. We will
prove this effect in Sec. IV-A (henceforth we set $K=1$). Note that the
potential energy in Eq.~(\ref{Eq1}) is periodic along $x$-axis at
fix $z$ independent of the chirality and $z$, but the
periodicity along $y$-axis depends on the chirality, i.e.
$E_T(x,y,z)\,=\,E_T(x+\mu\,d,y+\nu\,\delta_{m,n},z)$ where $\mu$ and
$\nu$ are two integer numbers and $\delta_{m,\,n}$ is the periodicity
number along $y$-axis, e.g. $\delta_{5,5}\cong\sqrt{3}a_0$
and $\delta_{9,0}\cong\,3\,a_0$. In fact, the periodicity of the ES is a consequence of the
crystalline structure and parallel alignment of the carbon
nanotubes. Moreover, the gas absorbtion into the CNTS depends on the ES periodic structure.

 The gradient of the potential is
the force experienced by the argon atom
\begin{equation}\label{Eq2}
\overrightarrow{\textbf{F}}\,=\,48\varepsilon{\sum^{N/2}_{i=-N/2}}{\sum^K_{t=0}}{\sum^M_{j=1}}{\sum^2_{l=1}}
\frac{(-1)^{l+2}\overrightarrow{\delta r}} {l\delta
r^2}(\frac{\sigma}{\delta r})^{12/l}.
\end{equation}
Newton's second law should be solved numerically for determining the
path of the motion.

\section{The continuum model}

Considering a hollow and continuous cylinder as a simple model for the
carbon nanotube~\cite{ACSnano} is the main assumption for the
continuum approach. Again, $N$+1 aligned tubes along $y$-axis are
separated by the distance $d$ in each layer of CNTS. The obvious difference between the
atomistic model and the continuum model is the absence of the
chirality effect in the continuum model. The other important
difference is the periodicity of ES along $y$-axis in the
atomistic model (not in the continuum model) which will be discussed
further in the next section. The LJ potential energy for $N$+1
cylinders which are interacting with a rare gas atom and are
parallel and aligned along the $y$-axis (when K\,=\,1) is
\begin{equation}\label{EqContinuum}
E_C(x,y,z)\,=\,\sum^{N/2}_{i=-N/2} E_i(x,y,z),
\end{equation}
where $E_i(x,y,z)$ is the contribution of the i$^{th}$ nanotube and is given by
\begin{equation}\label{Eq_i}
\small{E_i(x,y,z)\,=\,\sum^2_{l=1}\int^{2\pi}_{0}\int^{L}_{0}
\frac{2\,n_l(-1)^{l+1}\,R_{n,m}\,d\phi
dy_i}{[\Re^2_i-2\,Sin(\varphi_i)+(y-y_i)^2]^{6/l}}},
\end{equation}
with $n_l\,=\,{4\varepsilon\sigma^{12/l}}{\Sigma}$ where $\Sigma$ is the
mean surface density of carbon atoms for a graphene lattice which is
$\Sigma\,=\,\frac{2}{|\overrightarrow{a}_1\times
\overrightarrow{a}_2|}$. In Eq.~(\ref{Eq_i}) $(x_i,y_i,z_i)$ is the
coordinates of the chosen element from the i$^{th}$ nanotube, $L$ is an arbitrary length,
$R^2_{n,m}\,=\,{x^2_i}+{z^2_i}$ is the radius of each nanotube,
$\alpha_i\,=\,\tan^{-1}(\frac{x+i\,d}{z})$, $\varphi_i\,=\,\phi+\alpha_i$,
$\Re^2_i\,=\,R^2_{m,n}+R^2_i$, $R^2_i\,=\,(x+i\,d)^2+z^2$. Note that $``i"$ is
a positive integer number where $x_i\,<\,0$ and vise versa. In
Fig.~\ref{figmodel}(b) we illustrate the front view (in $x-z$ plane)
of only three nanotubes from the top layer. The relevant variables for the model are exhibited in
Fig.~\ref{figmodel}(b). The following integrals
are taken over the middle nanotube referring to $i$\,=\,1 in Eq.~(\ref{Eq_i}).

The integrals over the longitudinal viable ($y_i$ for the
i$^{th}$ cylinder) are done analytically and the integrals over the
azimuthal angle ($\phi$) are obtained numerically by employing
Mathematica software. The result of the integration over  $y_i$
for  repulsive term is
\begin{widetext}
\begin{equation}
I_{R}(x,\,y,\,z,\,\phi)=\int^{L}_{0}
\frac{n_1~dy_i}{\left(A+(y-y_i)^2\right)^{6}}\\ \nonumber
\end{equation}
\begin{equation}
=\frac{n_1 \left(\sqrt{A} \zeta  \left(965 A^4+2370 A^3 \zeta
^2+2688 A^2 \zeta^4+1470 A \zeta^6+315 \zeta ^8\right)+315
\left(A+\zeta^2\right)^5 \text{ArcTan}\left[\frac{\zeta
}{\sqrt{A}}\right]\right)}{1280 A^{11/2} \left(A+\zeta
^2\right)^5},\nonumber
\end{equation}
and for  attractive term is
\begin{equation}
I_{A}(x,\,y,\,z,\,\phi)=\int^{L}_{0}
\frac{-n_2~dy_i}{\left(A+(y-y_i)^2\right)^{3}}\nonumber
=-\frac{n_2\, \zeta \left(5 A+3 \zeta^2\right)}{8 A^2 \left(A+\zeta
^2\right)^2}-\frac{3\,n_2 \text{ArcTan}\left[\frac{\zeta
}{\sqrt{A}}\right]}{8 A^{5/2}},\nonumber
\end{equation}
where $A=\Re^2_i-2\,Sin(\varphi_i)$ and $\zeta=y-y_i$. Combining two
above integrals, the continuum model energy is written as
\begin{equation}
E_C(x,y,z)=\sum^{N/2}_{i=-N/2}
\int^{2\pi}_{0}2\,R_{n,m}\,(I_R(x,y,z,\phi)+I_A(x,y,z,\phi))|^{y_i=L}_{y_i=0}\,d\phi.\nonumber
\end{equation}
\end{widetext}

Differentiating the above equation gives the force experienced by
the argon atom due the interaction with i$^{th}$ tube:
\begin{equation}\label{Eq_fc}
\small{\overrightarrow{F_i}= \sum^2_{l=1}\int^{2\pi}_{0}\int^{L}_{0}
\frac{24 \overrightarrow{\Delta r}\,n_l(-1)^{l+1}\,R_{n,m}\,d\phi
dy_i}{l[\Re^2_i-2\,Sin(\varphi_i)+(y-y_i)^2]^{6/l+1}}},
\end{equation}
where $\overrightarrow{\Delta
r}=(x+i\,d-x_i)\hat{i}+(y-y_i)\hat{j}+(z-z_i)\hat{k}$.

 We will compare the results of the continuum model with those of
 the atomistic model in the next section.
\begin{figure*}
\begin{center}
\includegraphics[width=0.45\linewidth]{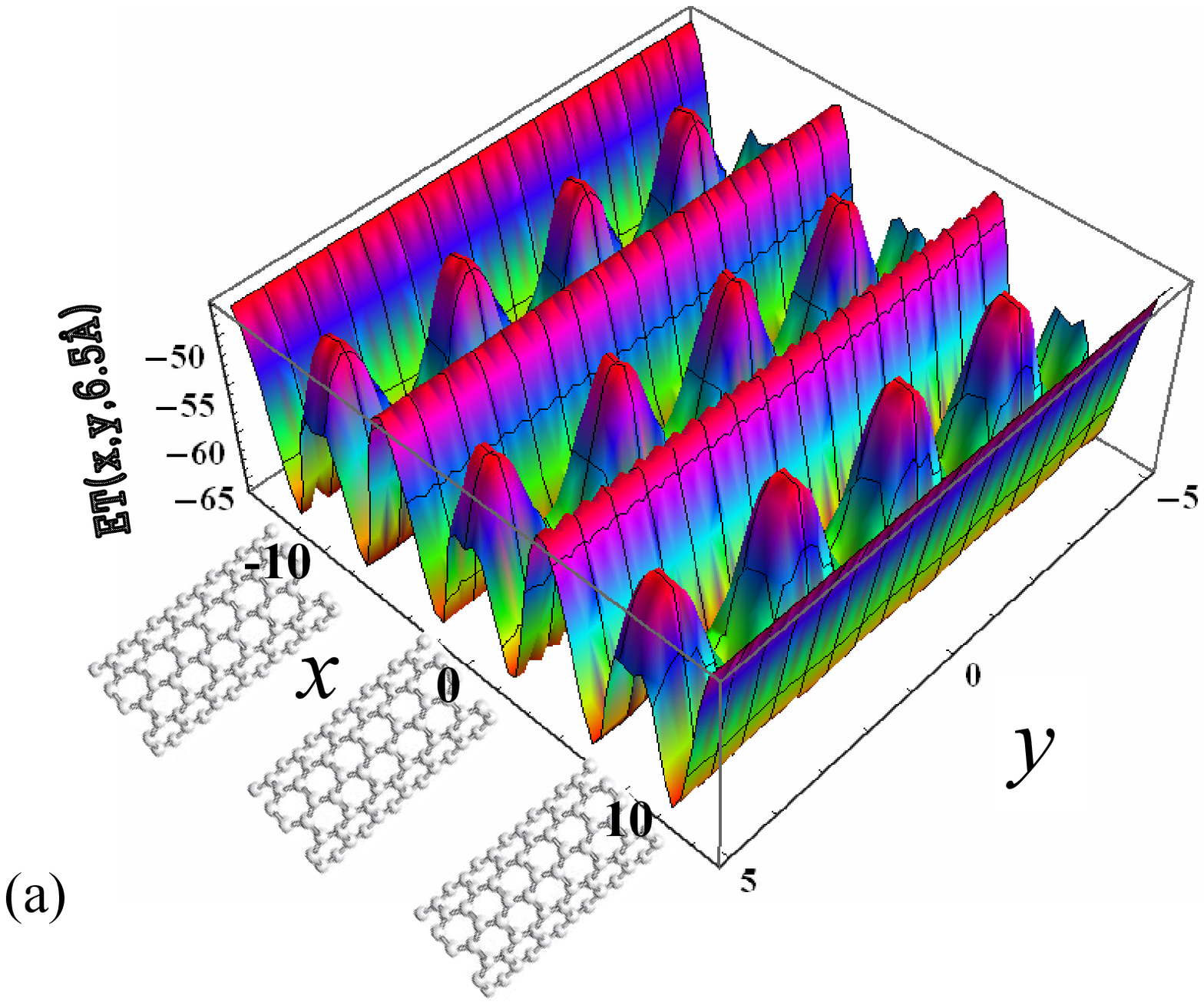}
\includegraphics[width=0.45\linewidth]{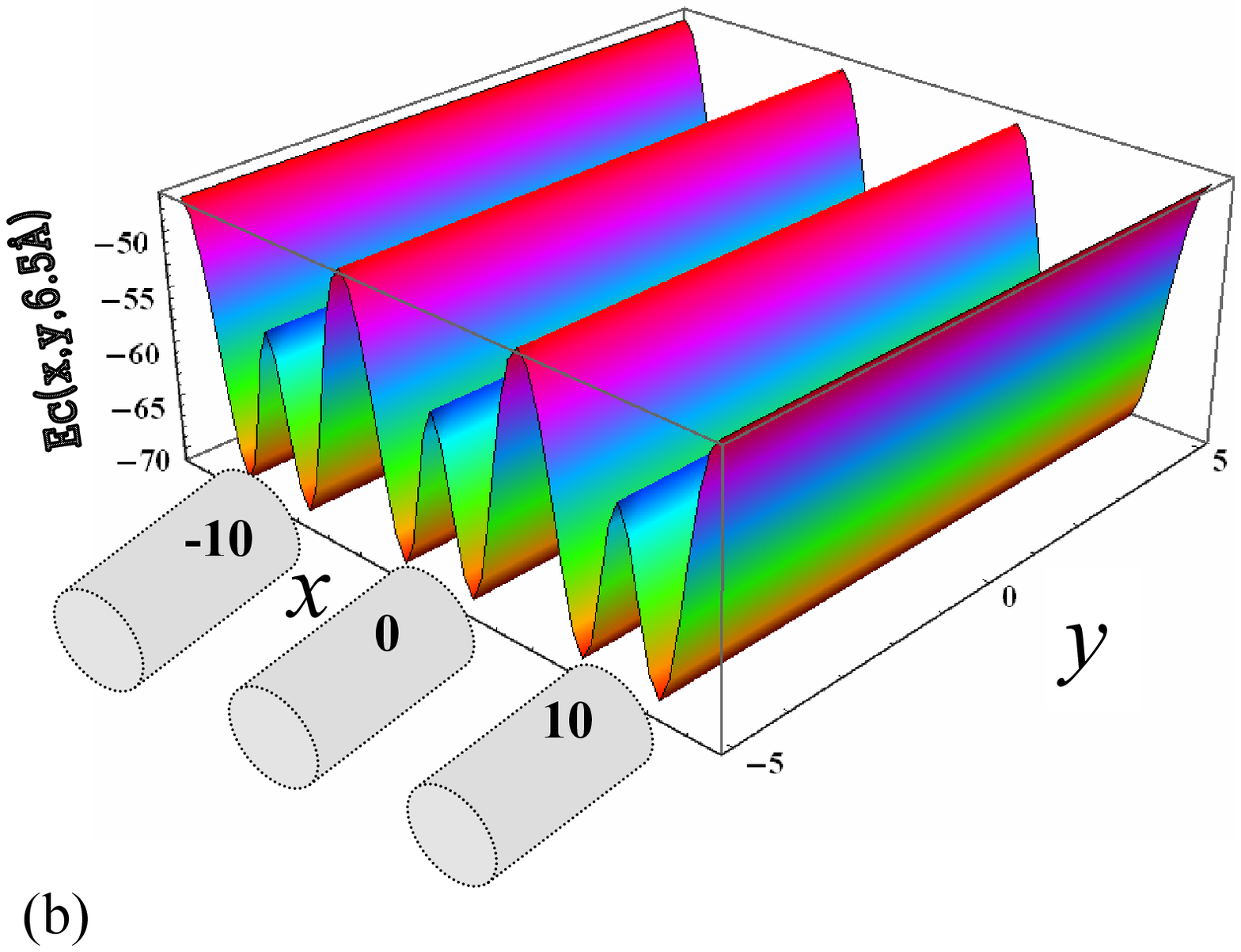}
\caption{(Color online) (a) The vdW energy surface of a CNTS
comprises of aligned armchair nanotubes interacting with
an argon atom (located at $z=6.5\,\AA$). (b) The vdW energy surface
of continuous tubes interacting with
an argon atom (located at $z=6.5\,\AA$). \label{figES} }
\end{center}
\end{figure*}

\begin{figure*}
\begin{center}
\includegraphics[width=0.45\linewidth]{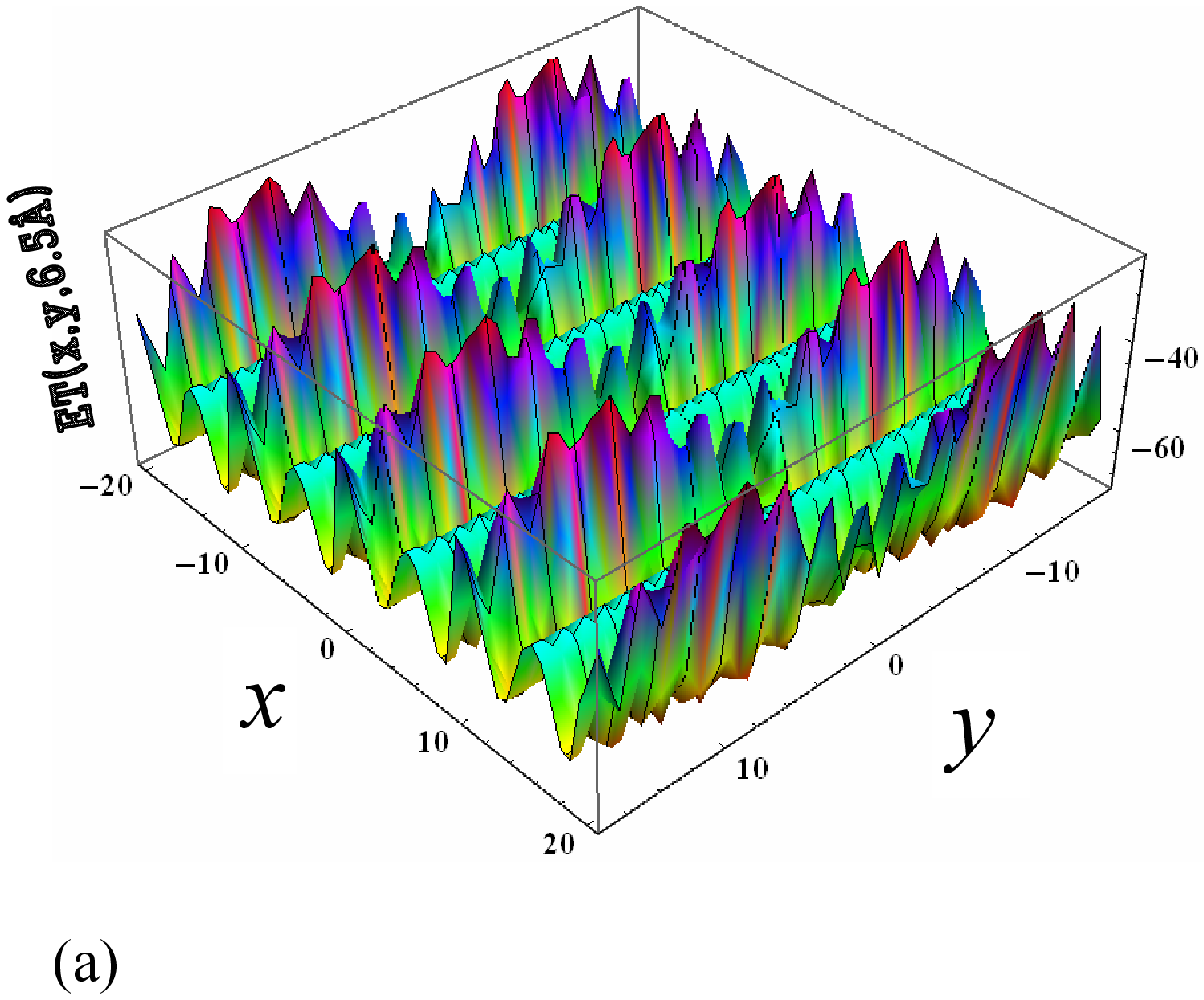}
\includegraphics[width=0.4\linewidth]{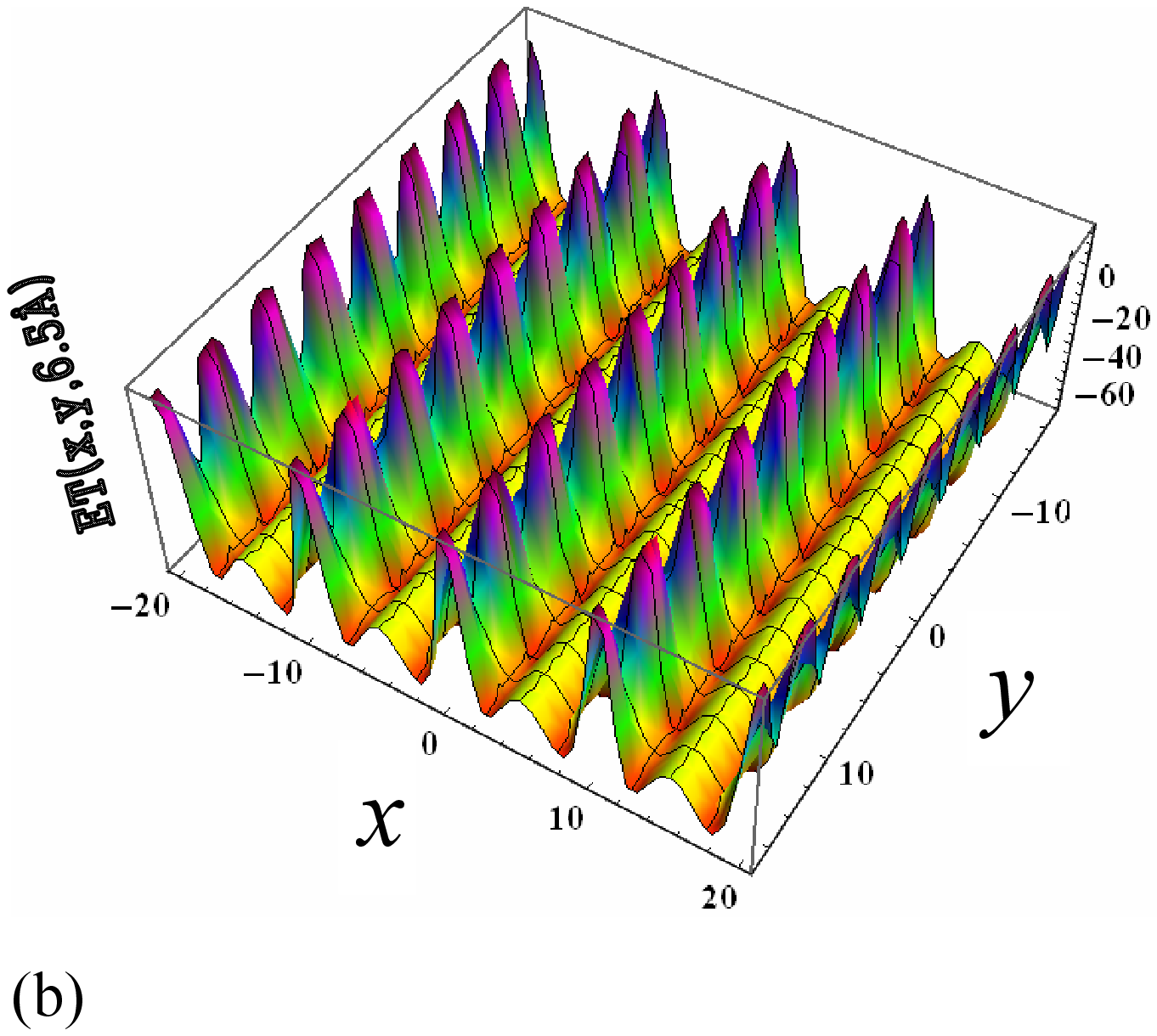}
\caption{(Color online) The vdW energy surface of chiral nanotubes with (6,4)
indexes in (a) and zigzag nanotubes with (9,0) indexes in (b), which are
interacting with an argon atom (located at $z$\,=\,6.5\,\AA).
\label{figES1} }
\end{center}
\end{figure*}

\section{Results and discussions}

\subsection{Energy surface}

Figure~\ref{figES} shows two energy surfaces (ES) in the $x-y$ plane above the CNTS at $z = 6.5\,\AA$ which
are related to the atomistic (Eq.~(\ref{Eq1}))
and the continuum (Eq.~(\ref{EqContinuum})) models, i.e. $E_T(x,y,6.5\,\AA)$
 and $E_C(x,y,6.5\,\AA)$ respectively.
 The ESs were calculated with $K$=1, $N$\,=\,7
and $d$\,=\,10\,\AA~(these are fixed in this study).
For the atomistic
model (Fig.~\ref{figES}(a)) we set $M\,=\,400$ for AC nanotubes with
(5,\,5) indexes.
The colored tubes indicate the location of the top
layer of the CNTS (includes eight aligned carbon nanotubes).
We found that the chosen values for $M$ and $N$ are
sufficiently large to eliminate boundaries when ES is considered in the range of
 $|x,y|\,<\,20\,\AA$.~In contrast to the ES obtained for
the graphene~\cite{neek}, from the ES in Fig.~\ref{figES}(a) one can not identifies
the honeycomb lattice structure of the nanotubes.
This is because of the curvature of the nanotubes. Moreover, the ES approaches to a flat
surface far away from the sheet.

As can be seen from Fig.~\ref{figES},
 the main difference between two models is the appearance of the periodic peaks along $y$-axis in the atomistic model.
 In spite of the continuum energy surface, i.e. $E_C$,
$E_T$ has many peaks/valleys along the $y$-axis (above $(i\,d,y)$ points for each integer `i').
These valleys make the connection between two adjacent channels
easier and  will also enhance the probability of the lateral motion (along $x$-axis) of the particle through the
mentioned valleys. Furthermore
the appearance of the peaks along the $x=i\,d/2$ line can be understood regarding to
the closed energy curves around a single nanotube (Fig.~\ref{figcontour}(a))
in the $x-z$ plane. Assume two such nanotubes get laterally closer from infinite distances.
At certain distances due to the overlap of their energy curves, the mentioned peaks appear.

We depict ES of ZZ-CNTS with (9,0) indexes and CR-CNTS with (6,4) indexes in
Fig.~\ref{figES1} (we used shorthand letters ZZ, AC and CR for ZZ-CNTS, AC-CNTS and CR-CNTS, respectively in
 the figures legends). Note that $R_{9,0}\approx R_{6,4}\cong
R_{5,5}\,=\,3.39\,\AA$~and that is the reason for choosing (9,0) and (6,4)
indexes. It is clear from Fig.~\ref{figES1} that the periodicity of
$E_T$ along $y$-axis depends on the $(m,n)$ indexes.

\begin{figure}
\begin{center}
\includegraphics[width=0.9\linewidth]{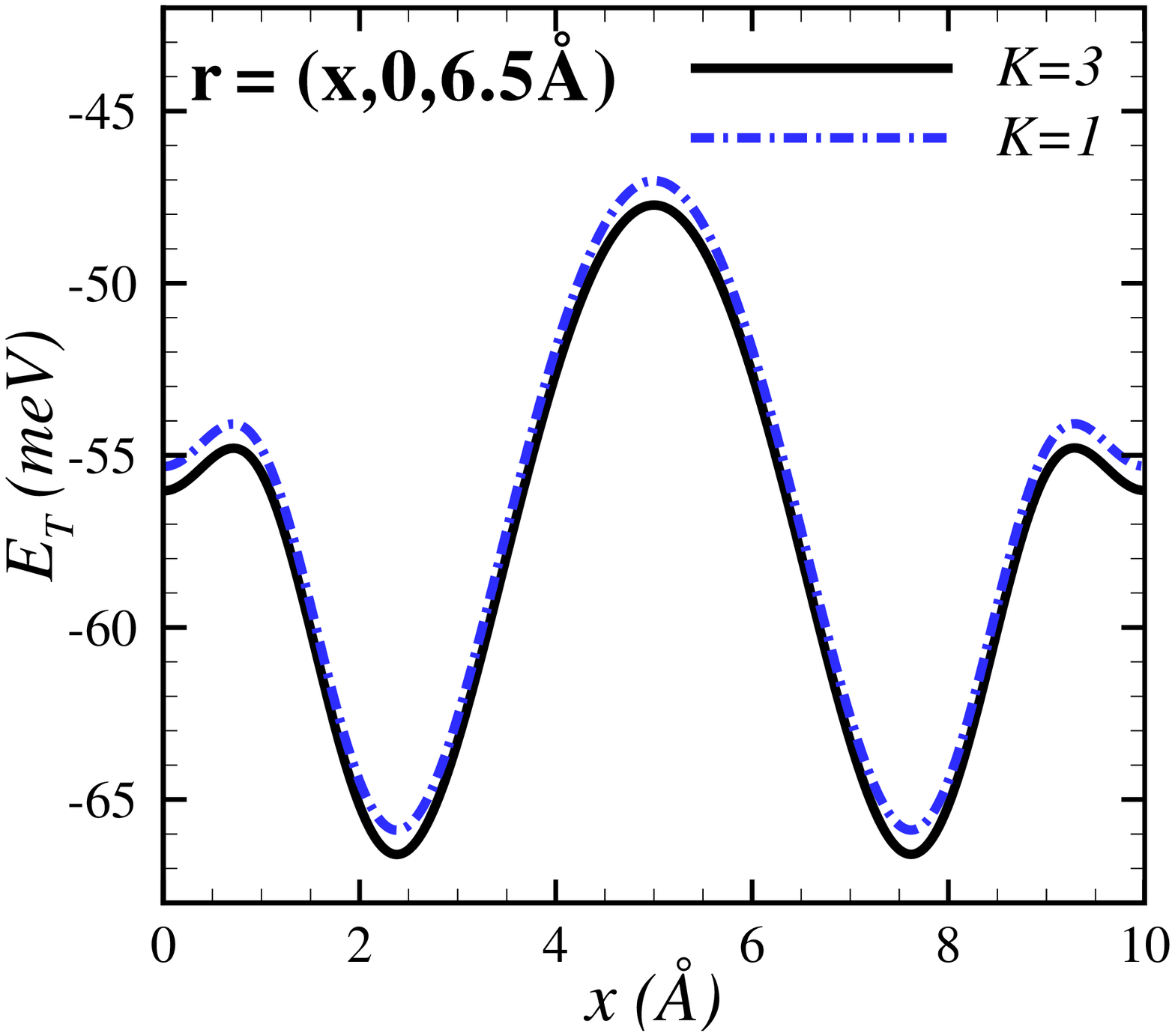}
\caption{(Color online) Comparison of the energy curves
 (as a function of $x$) produced by three layers of AC-CNTS (solid curve)
and by one layer of AC-CNTS (dashed curve) for
$y\,=\,0$ and $z\,=\,6.5\,\AA$. \label{k31}}
\end{center}
\end{figure}

Increasing the number of layers, i.e. $K$, varies  ES
slightly implying that the major part of the ES is due to the interaction with the top
layer. In order to prove this fact, we show the variation of
$E_T$ for AC-CNTS with $K$=\,3 and $K$=\,1  versus $z$ in Fig.~\ref{k31}.
The ratio $E_T|_{K\,=\,3}/E_T|_{K\,=\,1}$ varies around one, thus setting $K$\,=\,1 is a reasonable assumption.
 The same behaviors were obtained for the
variations of the mentioned ratio versus $x$ and $y$.

\begin{figure}
\begin{center}
\includegraphics[width=0.9\linewidth]{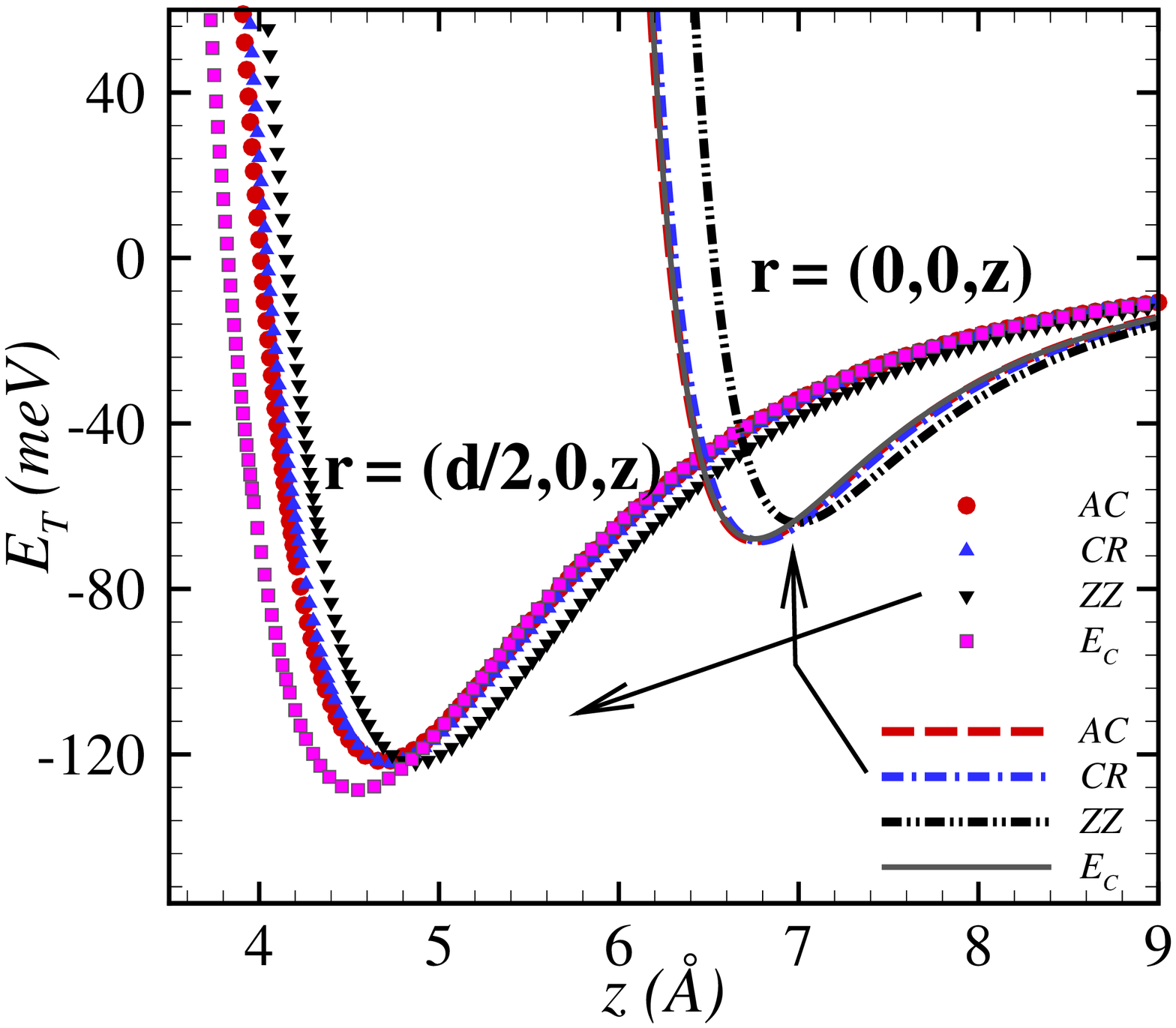}
\caption{(Color online) Variation of the energy versus $z$ for different type of carbon nanotube sheets above two points $(0,0)$ and
$(i\,d/2,0)$ in the $x-y$ plane. \label{figvz}}
\end{center}
\end{figure}

 Figure~\ref{figvz} shows the variation of ES for an AC-CNTS along $z$-axis at different ($x,y$)
 points, namely $E_T(0,0,z)$, $E_T(d/2,0,z)$ and $E_C(0,0,z)$, $E_C(d/2,0,z)$.
Although these curves are similar to
$u(r)$ function, but indeed they are different. The curves for ZZ-CNTS have
different minima as compared to those for AC-CNTS and CR-CNTS.
The minimum of energy curve for
ZZ-CNTS (above ($x\,=\,0,y\,=\,0$) point) is around $6.99\,\AA-R_{9,0}$ and is
about $6.75\,\AA-R_{5,5}$ for AC-CNTS and CR-CNTS.
The lowest energy occurs for $E_C$. In general, the energy curves above $(i\,d/2,y)$ points have a shift with respect
to those above $(i\,d,y)$, see Fig.~\ref{figvz}. The appearance of the
minima above $(i\,d/2,y)$ points are apparently in contrast to the appearance of the ES's peaks above the same points
in  Fig.~\ref{figES} and Fig.~\ref{figES1}.
In fact, this is an unpredictable effect that when $z$ decreases, the energy increases above $(i\,d,y)$
and decreases above the points between the tubes (i.e. $(i\,d/2,y)$)  independent of the $y$ values.
Above $(i\,d/2,y)$ points the minima are due to the free space between adjacent tubes, i.e.
the grooves between two adjacent nanotubes.
Therefore the
particles will prefer to aggregate between the tubes than above the tubes.
Notice that above $(i\,d/2,0)$ points the
binding energy (minima of the curves) is almost twice smaller
than the one above $(i\,d,0)$. The minima of the curves above $(0,0)$
 are around 70\,meV and above $(i\,d/2,0)$ points are about 128\,meV which
are 2.7 and 4.9 times larger than the thermal energy at room temperature, respectively.
The binding energy of an argon atom above a graphene sheet is around
$\simeq$\,85\,meV~\cite{Lajavardipour}, which is smaller
than the binding energy of an argon atom located at $(i\,d/2,0)$
above a CNTS.

\begin{figure*}
\begin{center}
\includegraphics[width=0.3\linewidth]{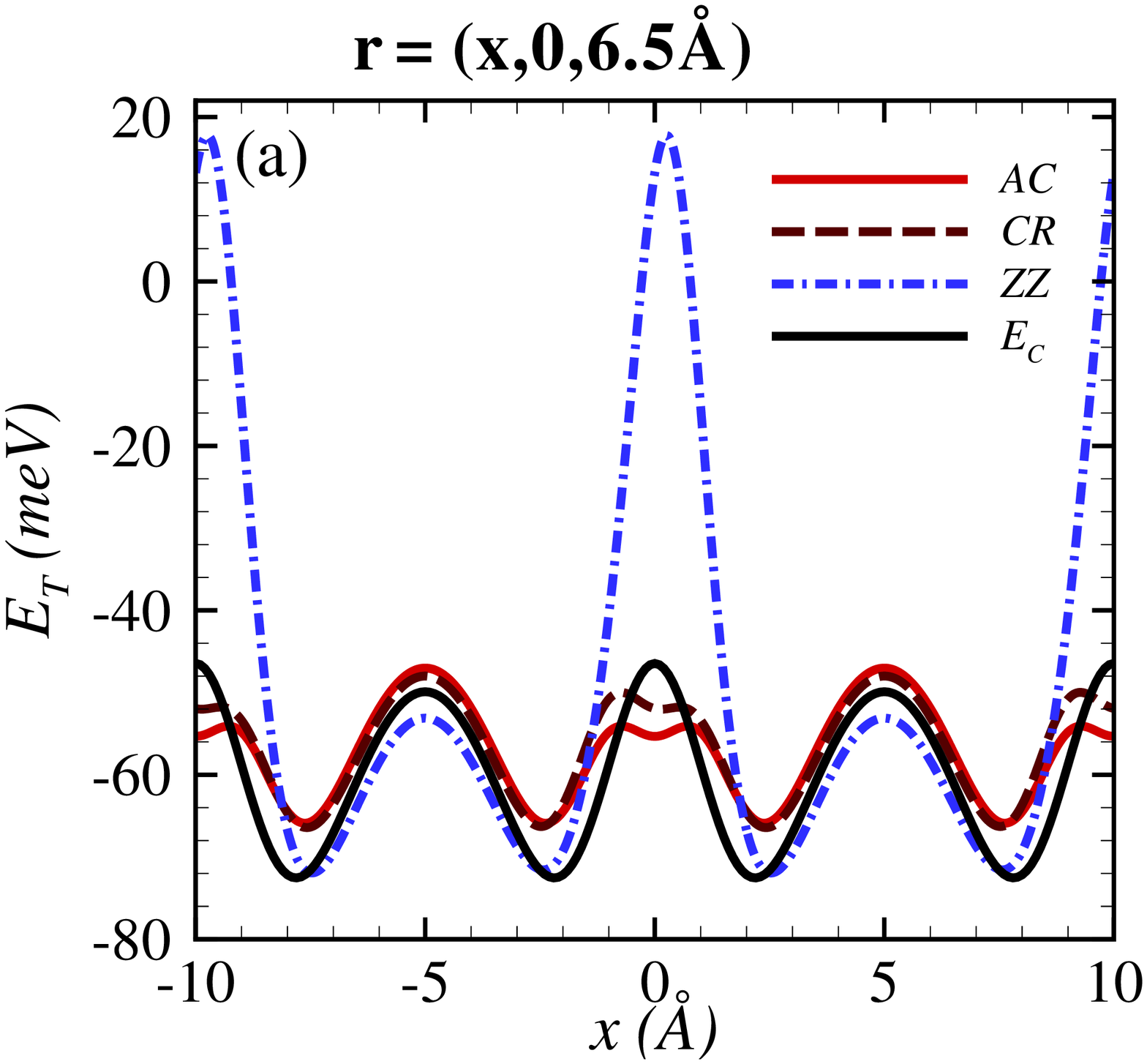}
\includegraphics[width=0.3\linewidth]{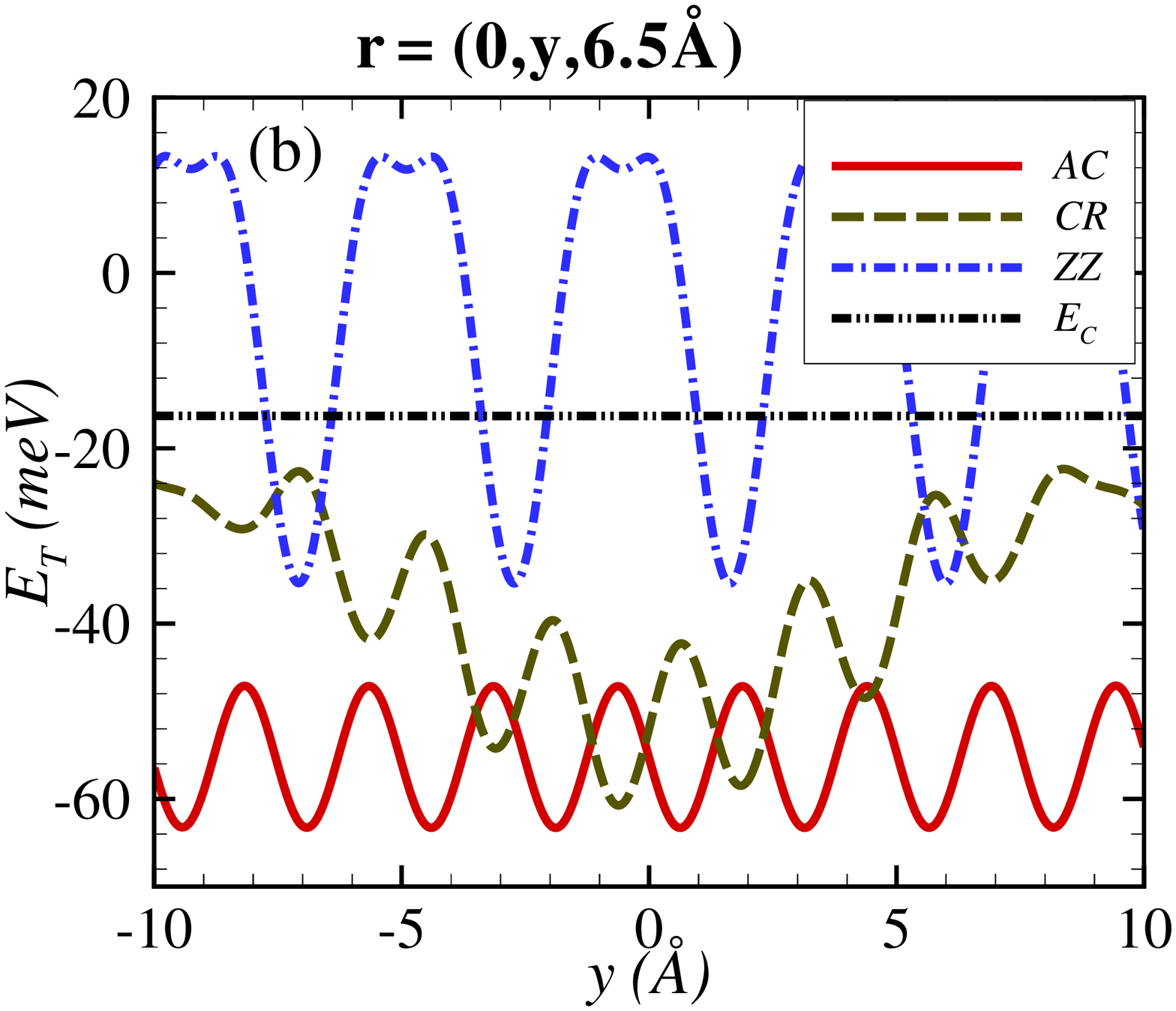}
\includegraphics[width=0.265\linewidth]{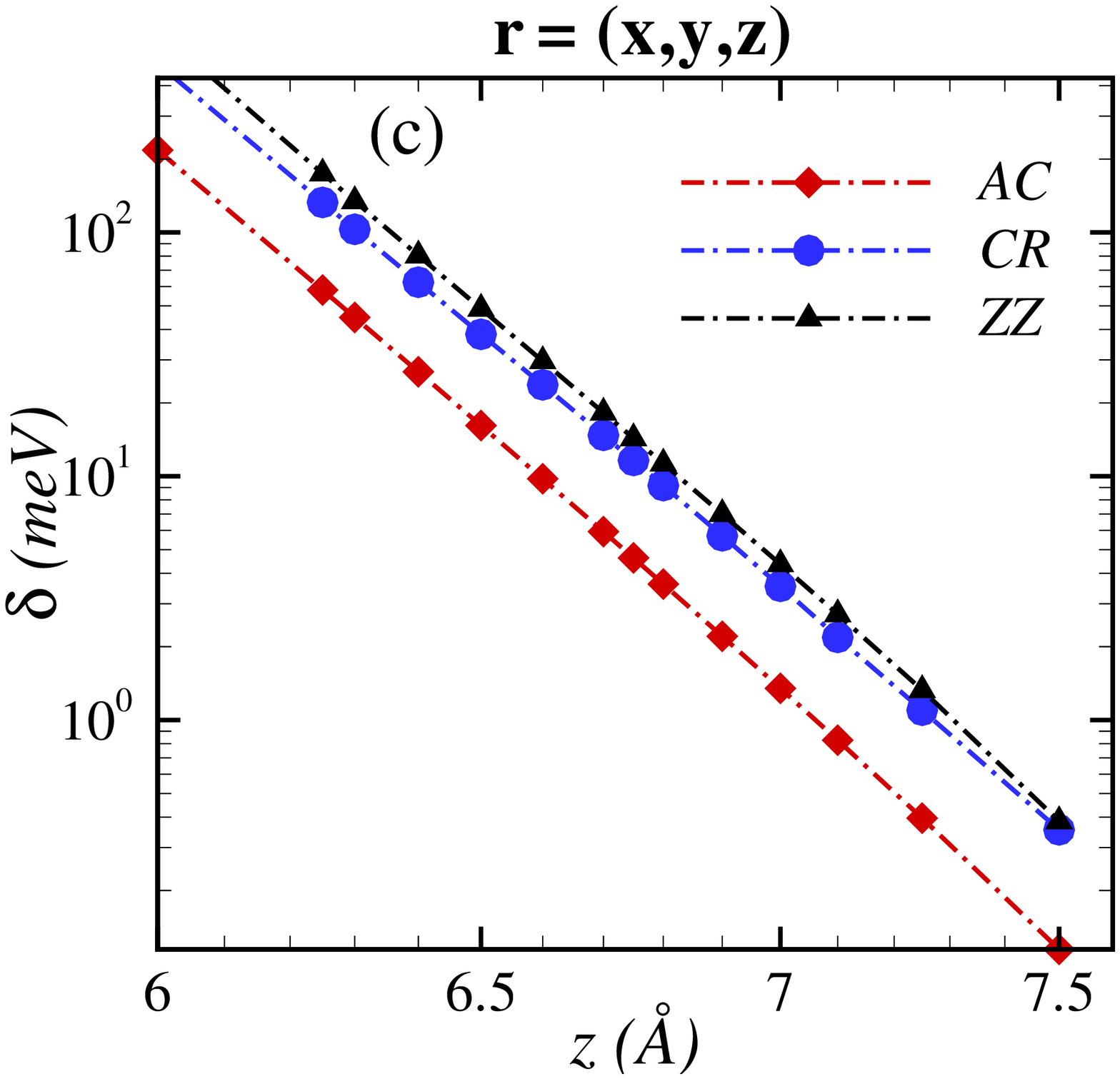}
 \caption{(Color online) The periodic vdW potential energy versus $x$ for ($x$,0,6.5\,\AA)~in (a) and versus $y$
 for (0,$y$,6.5\,\AA) in (b). (c) Variation of the
energy barriers ($\delta\,=\,E_{max}(x,y,z)-E_{min}(x,y,z)$) along $z$.
\label{figxy}}
\end{center}
\end{figure*}

\begin{figure*}
\begin{center}
\includegraphics[width=0.3\linewidth]{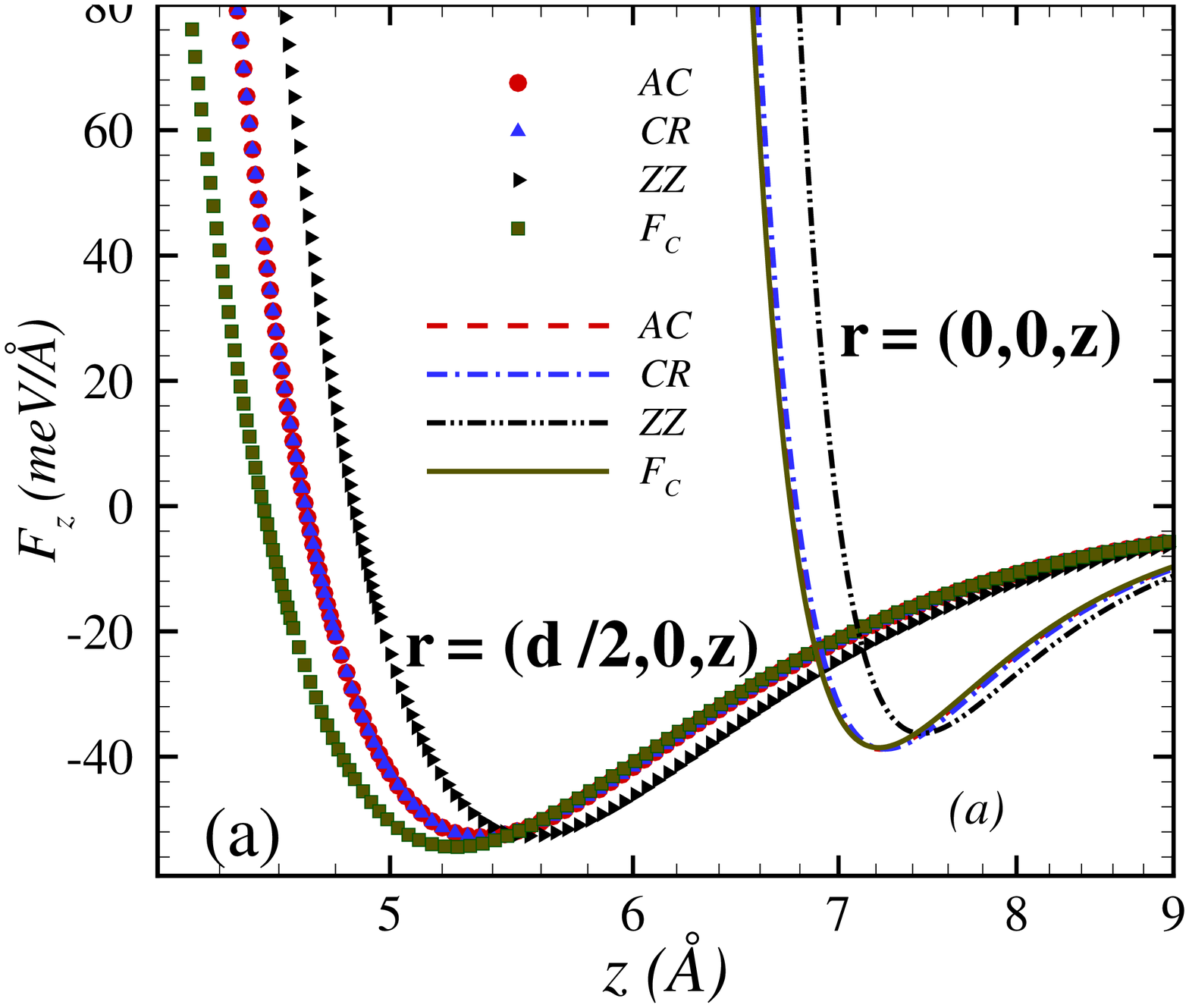}
\includegraphics[width=0.3\linewidth]{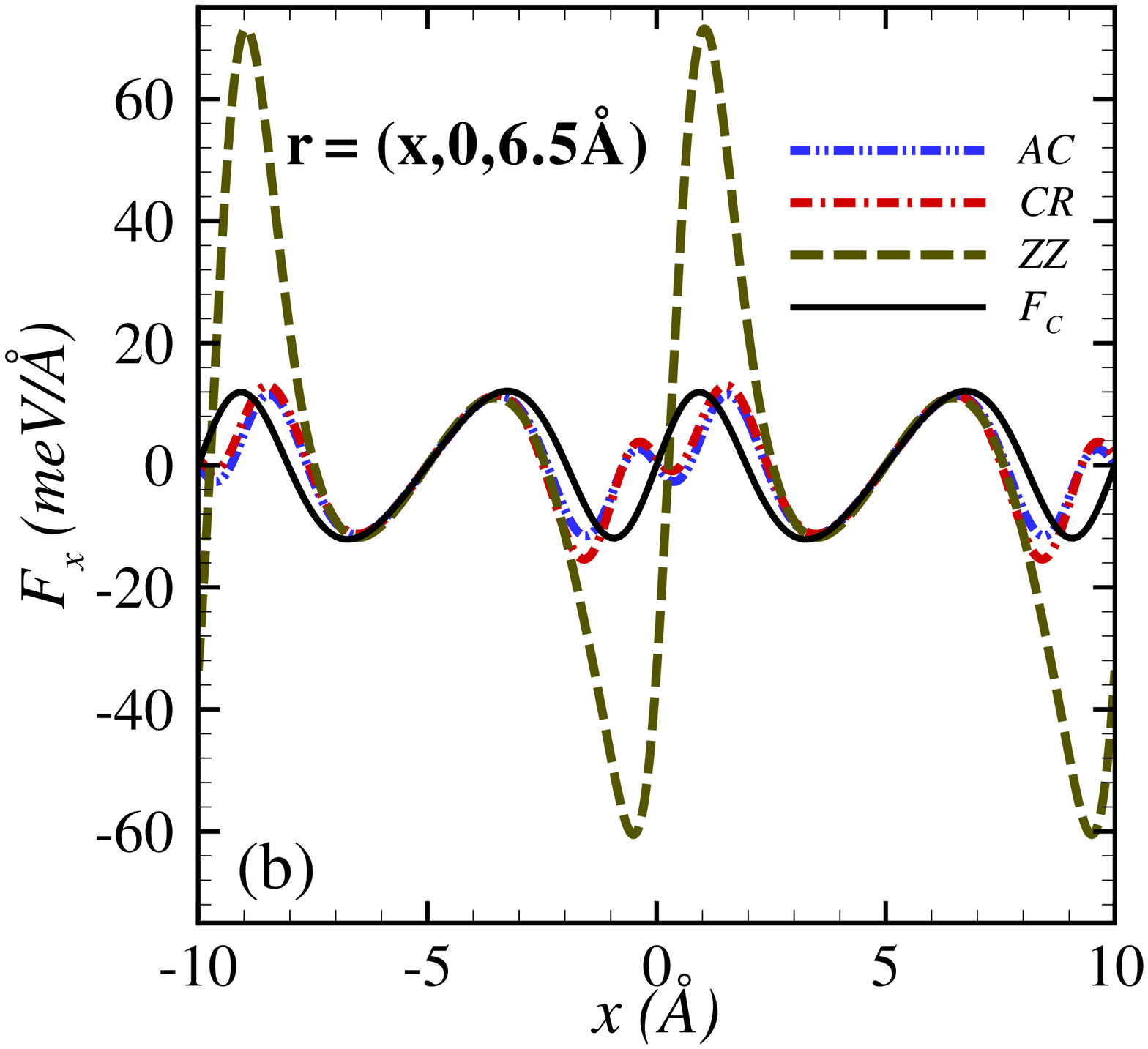}
\includegraphics[width=0.3\linewidth]{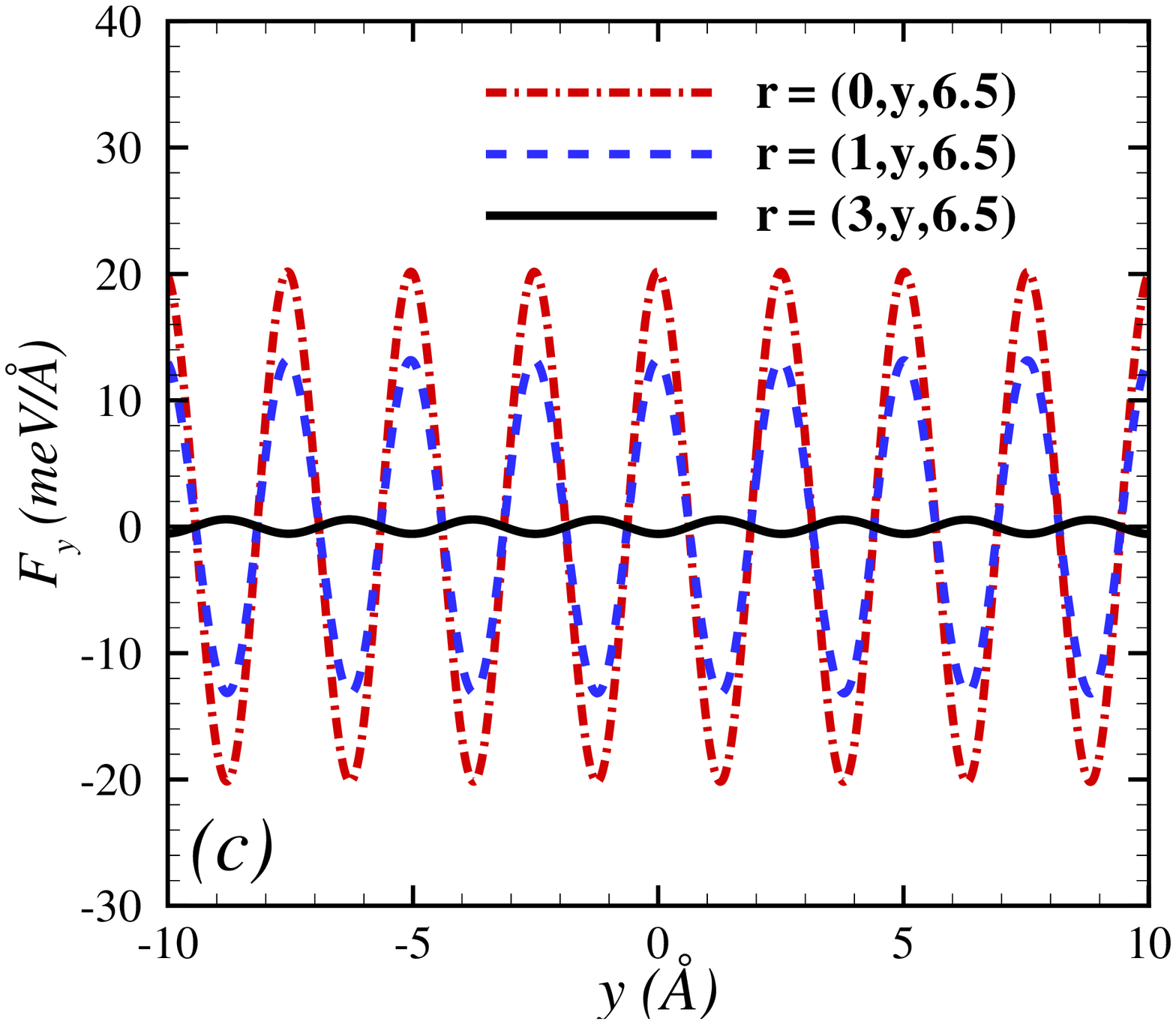}
\caption{(Color online) (a) Variation of the $z$-component of the
force versus $z$ which is acting on an argon atom located above (0,0) point.
(b) Variation of the  $x$-component of the force versus $x$ which is acting on an argon atom located at
 $(x,0,6.5\,\AA)$.~ (c) Variation of the $y$-component of the force versus $y$ which is acting on an argon atom located
at $(0,y,6.5\,\AA)$, $(1,y,6.5\,\AA)$ and $(3,y,6.5\,\AA)$. \label{figF}}
\end{center}
\end{figure*}

To investigate the periodicity of energy along $x$ and $y$-axis the
variations of ES are shown in Fig.\ref{figxy}. Notice that the
variations of energy versus $x$ are almost the same for AC, CR
and $E_C$ while they are different from ZZ. In all cases the minima of
the curves are equal, independent of the chirality an $z$ values.
Figure~\ref{figxy}(b) shows the variations of energy versus $y$
at $z=6.5\,\AA$. In spite of the curves in Fig.~\ref{figxy}(a),
the periodicity along $y$ depends on the chirality. The $E_C$ is a constant along $y$-axis.

The maximum barriers
are found for ZZ-CNTS and the minimum are for AC-CNTS. The barriers in the CR-CNTS
vary with $y$ resulting the possibility of the absorbtion of external atoms in
different energy levels. However, the barriers in the AC-CNTS are constant for fixed $z$,
which are consequence of the sine shape of the curves.
Figure~\ref{figxy}(c) shows the variation of the energy barriers ($\delta\,=\,E_{max}-E_{min}$)
versus $z$ in log-log scales. The barriers decrease
non-linearly with $z$ which implies that far from the sheet, the particle has free-like motion.

\subsection{Force calculations}

Using Eq.~(\ref{Eq2}( and Eq.~(\ref{Eq_fc}) we calculated the various components
of the force experienced by an atom above CNTSs with different chiralities. Here we
report the variation of $F_z$ versus $z$, $F_x$ versus $x$ and $F_y$ versus $y$.
Figure~\ref{figF}(a) shows the variation of the $F_z$  versus $z$ at two points (0,\,0) and
($d$/2,\,0). Clearly the ZZ-CNTS attracts particle
differently. The forces on the argon atom at ($i\,d$/2,0) point is more attractive than the one at $(i\,d$,0)
implying tendency of the particles to aggregate between the tubes.

Figure~\ref{figF}(b) shows the variation of the
$F_x$ versus $x$ for various $y$ at
$z\,=\,6.5\AA$. As we see the curves for the AC-CNTS and the CR-CNTS are similar.
 Figure~\ref{figF}(c) shows the variation
of the $F_y$ for an AC-CNTS versus $y$ for
different $x$ and  $z\,=\,6.5\AA$.  The absolute values of the forces
decrease with increasing $x$.

\subsection{Equations of motion}
\begin{figure*}
\begin{center}
\includegraphics[width=0.3\linewidth]{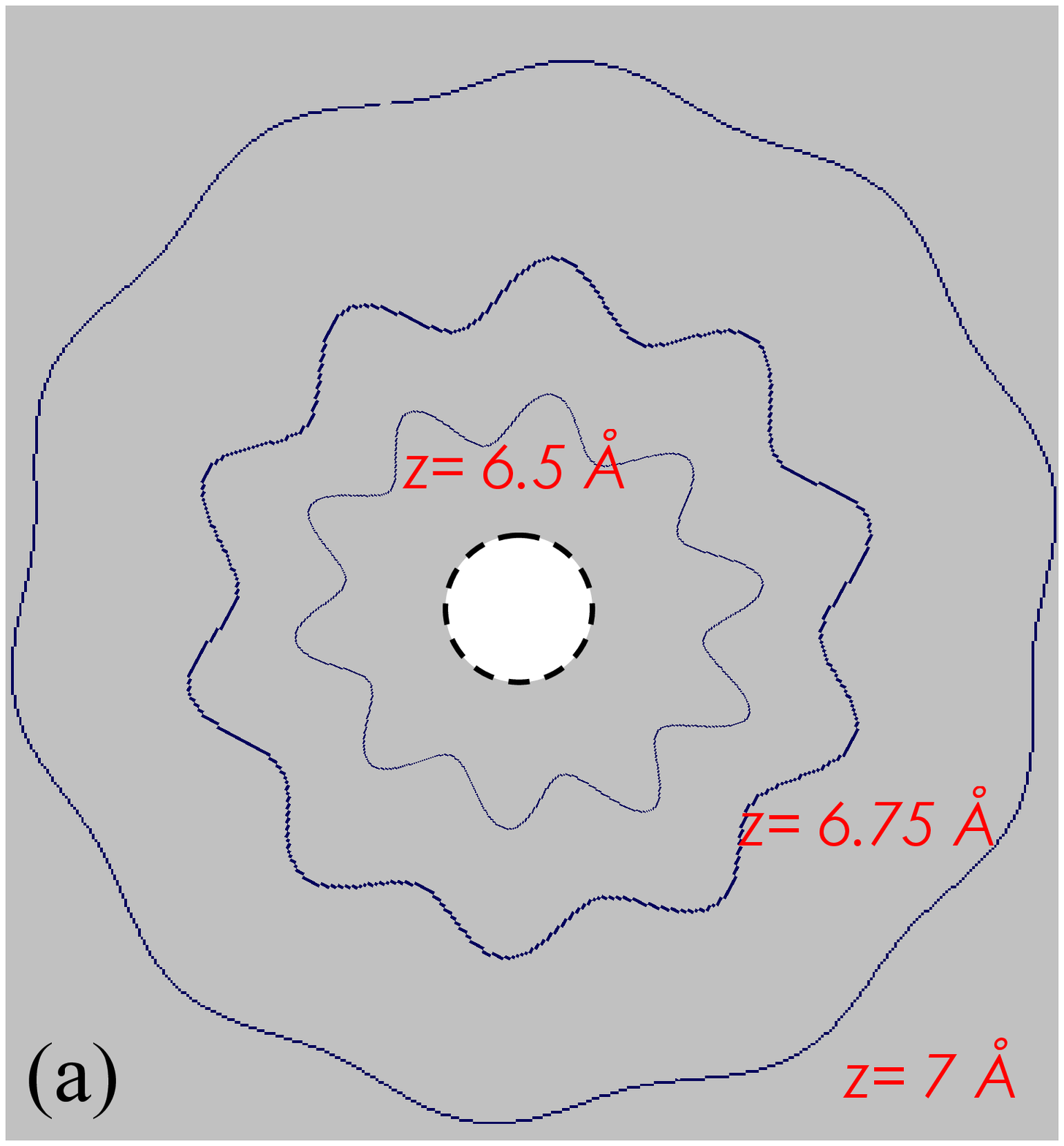}
\includegraphics[width=0.25\linewidth]{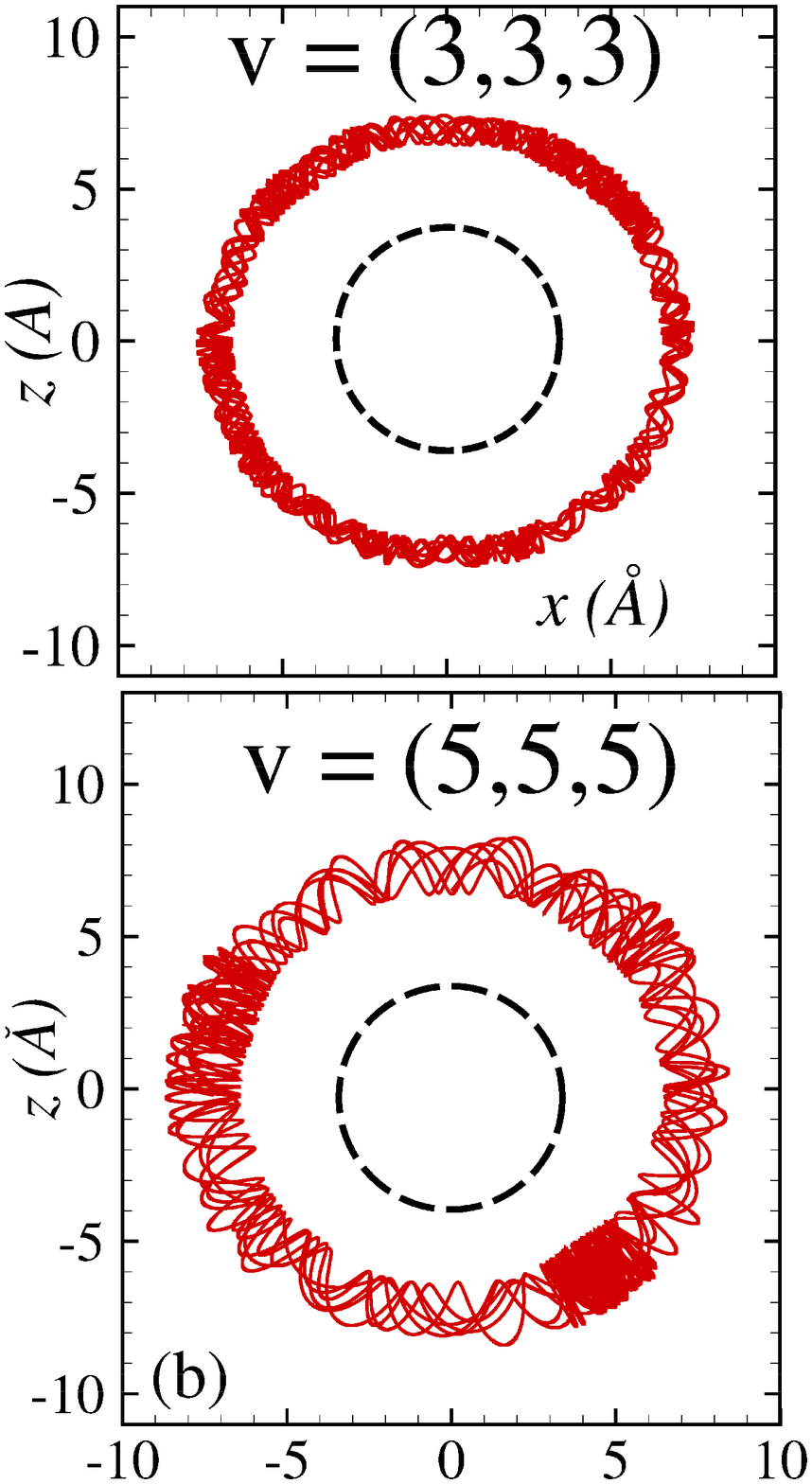}
\includegraphics[width=0.35\linewidth]{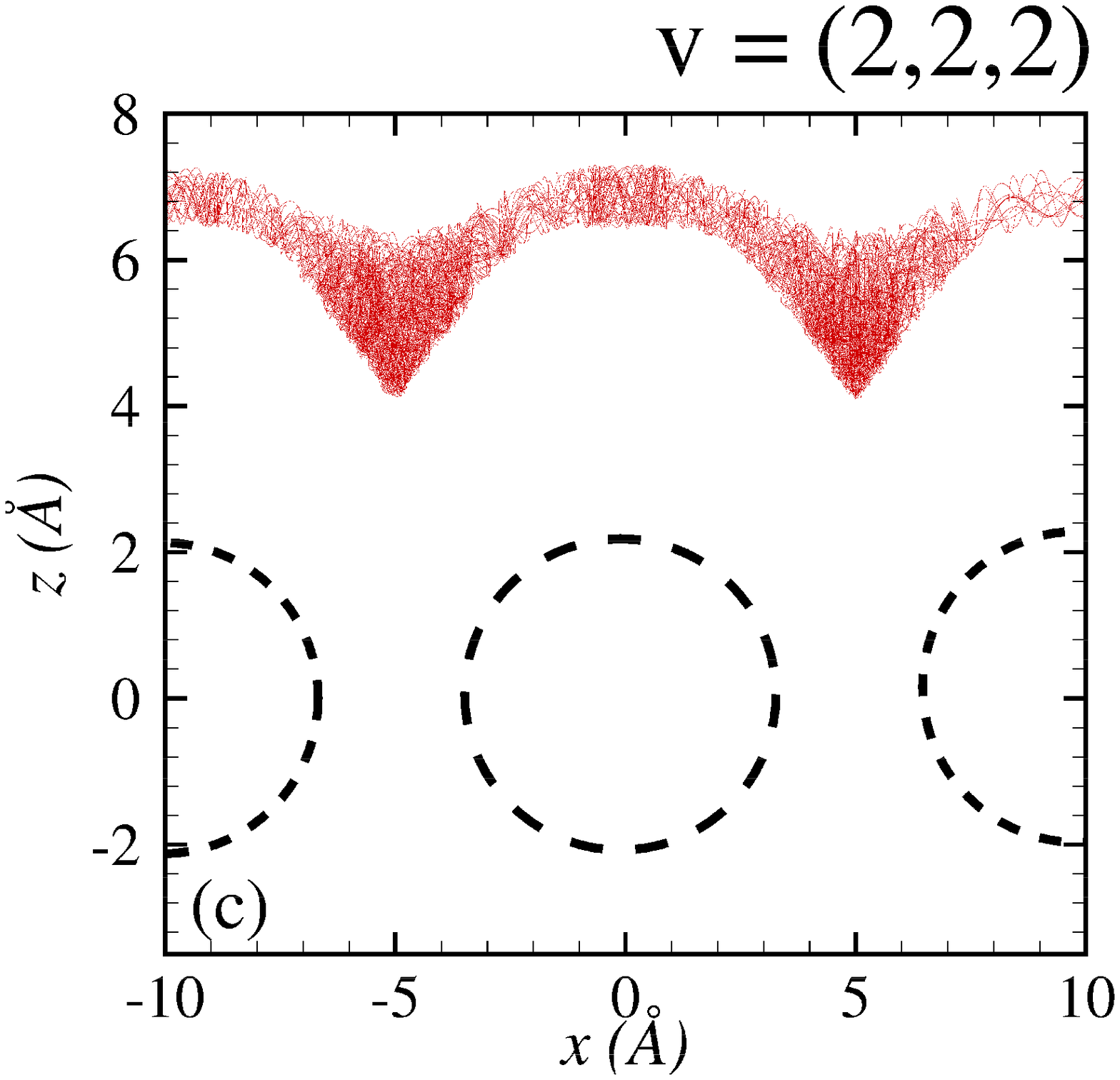}
\caption{(Color online) (a) Front view of the energy curves at
various heights (b) and the path of the motion of an argon atom
(with two different initial velocities) around an armchair
nanotube. (c) The path of the motion of an argon atom with initial
velocity (2,2,2) in semispace above a CNTS. \label{figcontour}}
\end{center}
\end{figure*}

In order to find the path of the motion of an argon atom above the CNTS,
 we solved equations of motion numerically. The
aim is to find  the influence of the ES on the path of motion of an external
atom. The study of an argon atom close to a single nanotube surface provides further
 insights into the problem. Therefore it would be helpful to show how ES energy curves varies
  around an individual carbon nanotube. Figure~\ref{figcontour}(a) shows the energy
curves for three $z$ values, i.e. $z$=\,6.25\,\AA,~6.5\,\AA,~6.75\,\AA,~7\,\AA. These cycloidal closed curves have been obtained by transforming
 the Cartesian coordinates (in Eq.~(\ref{Eq1}))
 to the polar coordinates. It is observed that with increasing $z$, the curves close to the circular shape which
 results in the flattening of ES at distances far away from the sheet

The numerical solutions is presented for both the motion of an argon atom around an individual
AC carbon nanotube and an AC-CNT. The velocity Verlet algorithm is used for the
numerical integration and the time is step set to be 1\,fs. To solve the
Newoton's equation one needs two initial conditions, namely the
initial position of the argon atom and its initial velocity. We set
the initial position to (0,0,6.5\,\AA)~ and the velocity vector to an
arbitrary vector $\overrightarrow{v}=(v_x,v_y,v_z)$. We found that
the larger initial velocities (corresponds to the higher kinetic
energy of the atom) give the larger probability for passing the
barriers. Figure~\ref{figcontour}(b) shows the front view of the two
typical paths with $\overrightarrow{v}=(3,3,3)$ and
$\overrightarrow{v}=(5,5,5)$. It is also found that argon atoms with larger initial
 velocity, have larger fluctuation amplitude around the tubes.

In addition, we give the results for the path of the motion of an argon atom
above an AC-CNTS. Figure~\ref{figcontour}(c) shows the front view of the
resultant. Notice that the particle prefer to aggregate within the grooves.
Increasing the initial velocity of the particle enables the
particle to overcome the barriers and visit more points at higher heights.
In Fig.~\ref{figcontour}(b) and Figure~\ref{figcontour}(c) the dashed inner circles indicate the
carbon nanotubes locations.

\subsection{Defected CNTS}

\begin{figure*}
\begin{center}
\includegraphics[width=0.50\linewidth]{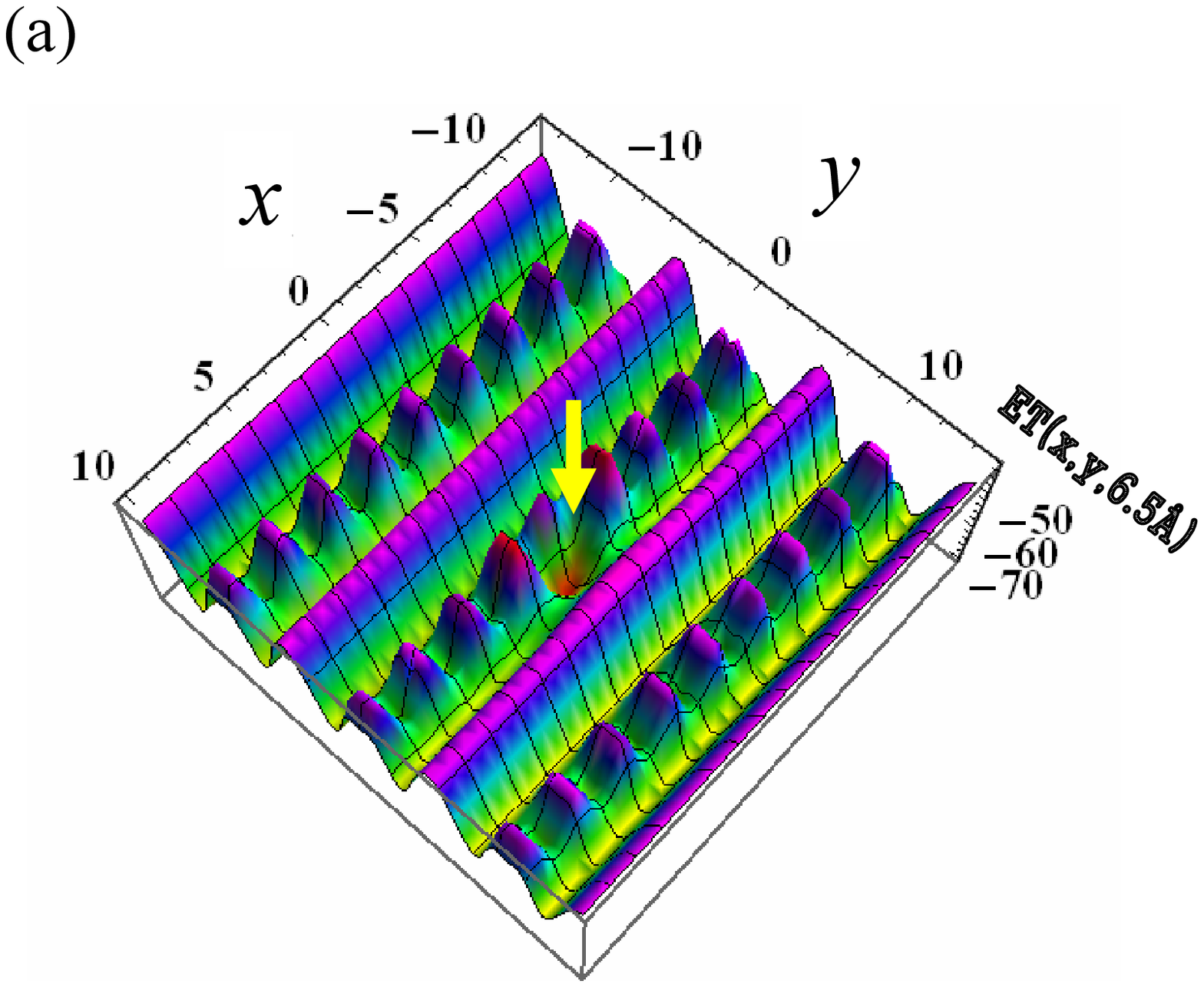}
\includegraphics[width=0.40\linewidth]{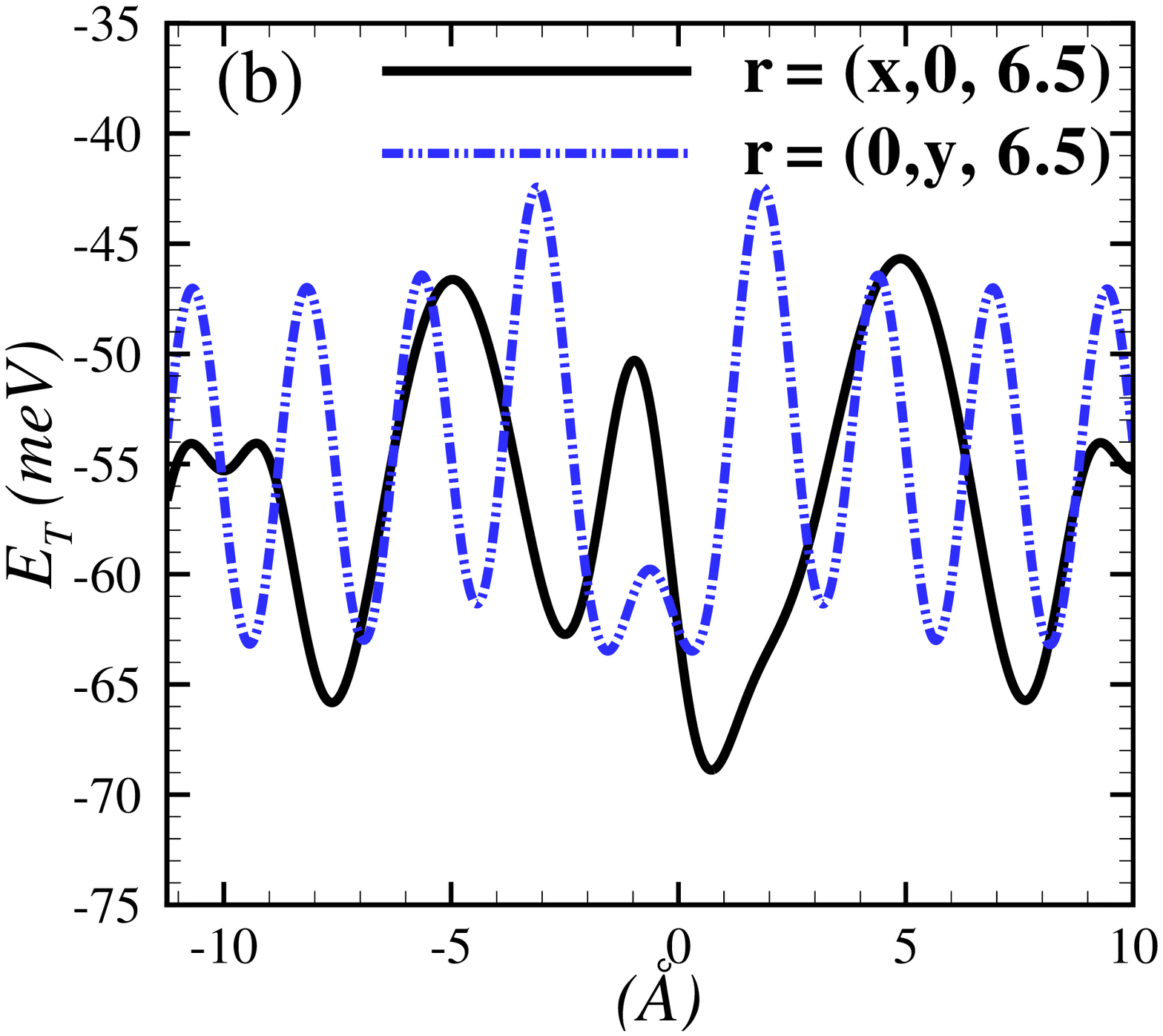}
\caption{(Color online) (a) Energy surface for an AC CNTS with a
vacancy in the center of the middle tube. The yellow vertical arrow indicates
the  vacant site. (b) The variation of the energy
versus $x$ and $y$, respectively. For the dashed curve
$x$\,=\,0 and for the solid curve $y$\,=\,0 for $z\,=\,6.5\,\AA$.
\label{figdefect}}
\end{center}
\end{figure*}

Removing an atom from a perfect carbon nanotube makes CNTS defective. The
ES varies in the presence of the defect. Defects changed
the absorbtion mechanism and the paths of the motion of external atoms.
Figure~\ref{figdefect} shows the ES of an AC-CNTS (with (5,5)
indexes) where vacant site was put in the middle of the central nanotube. Notice that the
periodicity of the ES is altered around the vacancy.
 Vertical yellow arrow refers to the vacant site. We show the variation of $E_T(x,y,z)$
  versus both $x$ and
$y$ in Fig.~\ref{figdefect}(b). The solid curve corresponds to
the variation versus $x$ and the dashed curve is related to the variation versus $y$.
Comparing these curves with those shown in Fig.~\ref{figxy}(b,c) for perfect CNTS
implies that the energy barriers are
reduced due to the presence of the defect (gives more free motion along
the $x$-direction). We have performed the same calculations for the CNTS
with different chirality and found similar behaviors. Moreover, solving equations of motion
in this case showed that the regular aggregation between and above the tubes is missed.
Increasing the number of vacant sites causes aperiodic energy surfaces.

\section{Conclusions}
Both the atomistic and the continuum models were employed to study
the van der Waals energy surface around a carbon nanotube sheet. The
continuum model can not reveal the atomistic features of the
problem. The carbon nanotube sheet comprises zigzag carbon
nanotubes, shows different periodicity in the energy surface as
compared to those obtained from armchair and chiral carbon nanotube
sheets. The equation of motion was solved numerically and the path
of the motion of a rare gas atom moving within the energy surface
was determined. The continuum model predicts the constant energy
along tubes axes which is in contrast to the results obtained from
the atomistic model. The energy barriers decrease rapidly with
height. Defects disturb the periodicity of the energy surface,
reduce the barriers and change the gas absorbtion mechanism.

\end{document}